\documentclass[aps,amssymb,twocolumn,reprint]{revtex4}

\usepackage{graphicx}
\usepackage{psfrag}
\usepackage{color}
\usepackage{amsbsy}
\usepackage{amssymb}

\def\ve{\varepsilon}

\begin{document}


\title{A length-dynamic Tonks gas theory of  histone isotherms}

\author{Tom Chou}
\affiliation{Dept. of Biomathematics and Institute for Pure \& Applied Mathematics, UCLA, Los Angeles, CA 90095}



\date{\today}

\begin{abstract} 

We find exact solutions to a new one-dimensional (1D) interacting particle theory and apply the results to the
adsorption and wrapping of polymers (such as DNA) around protein particles (such as histones).  Each adsorbed
protein is represented by a Tonks gas particle.  The length of each particle is a degree of freedom that
represents the degree of DNA wrapping around each histone.  Thermodynamic quantities are computed as functions of
wrapping energy, adsorbed histone density, and bulk histone concentration (or chemical potential); their
experimental signatures are also discussed. Histone density is found to undergo a two-stage adsorption process as
a function of chemical potential, while the mean coverage by high affinity proteins exhibits a maximum as a
function of the chemical potential. However, {\it fluctuations} in the coverage are concurrently maximal. 
Histone-histone correlation functions are also computed and exhibit rich two length scale behavior.  

\end{abstract}

\maketitle


\vspace{3mm}

\begin{figure}
\begin{center}
\includegraphics[height=1.6in]{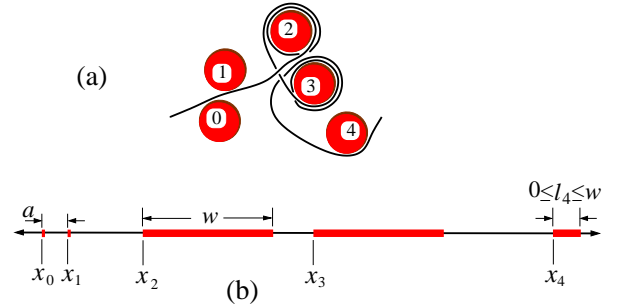}
\end{center}
\caption{(a) A histone-covered segment of DNA.
DNA can partially wrap nucleosome particles 
depending on solution ionic strength \cite{WIDOM2,NETZ}.
(b) The 1D representation of the 
footprint of the histone-wrapped regions of DNA (thick sections).
The associated particle lengths can vary from $a$ to $a+w$.}
\label{HISTONE}
\end{figure}

One-dimensional theories in statistical mechanics have been successfully applied to numerous
biophysical systems, including DNA denaturation \cite{PEYRARD}, particle transport across biological
channels \cite{CHOU,CHOUWATER}, adsorption on 1D substrates \cite{DIAMANT,VILFAN}, and
transport \cite{FREY} along microtubules. In this Letter, we study protein-DNA binding and wrapping
by solving a new 1D theory of interacting particles with dynamically varying particle sizes. DNA-histone
protein complexes (nucleosomes) play vital roles in compacting DNA and regulating nucleic acid
processing by mediating the accessibility by other regulatory proteins \cite{SINDEN,WIDOM,UCLA}. 
As shown in Fig.  \ref{HISTONE}a, DNA can wrap around each histone protein complex at most 1.7
times (about 146 base pairs) \cite{XRAY1,WIDOM2}.  The base pairs that come into contact with each
histone protein defines a footprint which we associate with particle lengths.  We consider the
collective behavior of the particles mediated only by the mutual exclusion of their footprints along a
tensionless DNA substrate. We neglect the statistical mechanics of the linker DNA between the histone
particles as well as the non-nearest-neighbor nucleosome-nucleosome interactions arising from their
relative, three-dimensional conformations \cite{RUDNICK}.

We now derive a generalization of the Tonks-Takahashi gas \cite{TONKS,THOMPSON}  
and apply it to DNA-histone structure.  Consider a 1D collection of $N$ particles
labeled  by the positions $x_{i}$ of their left-most edges and confined within length $L$.  The
first particle is fixed at $x_{0}=0$.  The minimum size $a$ of each particle defines the infinite
hard core repulsive interaction between adjacent particles such 
that $\vert x_{i}-x_{i\pm 1} \vert \geq a$ and $Na \leq L$.  Each 
particle $i$ also carries an additional internal degree of freedom
$\ell_{i}$ which corresponds to its length.  The length $\ell_{i}$ may, as we shall see, describe
the footprint of histone $i$ on the DNA substrate (Fig.  \ref{HISTONE}a).  
Since thermodynamic  properties depend only on the
interactions among particles, we compute the 
configurational partition function 

\begin{equation}
\begin{array}{l}
\displaystyle Z(N,L) = \int_{0}^{L_{0}}dy_{N-1}\cdots\int_{0}^{y2}dy_{1}
\prod_{i=0}^{N-1}f(y_{i+1},y_{i}),
\end{array}
\label{Z0}
\end{equation}

\noindent where  $y_{i} \equiv x_{i}-ia$, $y_{N}\equiv L_{0} \equiv L-Na$, and  

\begin{equation}
f(y_{i+1},y_{i}) = \int_{0}^{\ell_{i}^{*}} 
d\ell_{i}\exp\left[-\int_{0}^{\ell_{i}}\ve(x_{i}+\ell')d\ell'\right],
\label{F}
\end{equation}

\noindent where $\ell_{i}^{*}=\ell_{i}^{*}(y_{i},y_{i+1})$ is the maximum possible wrapping around particle $i$
and $\ve$ is a mesoscopic free energy for unit length extension. This winding-in energy consists of the molecular
binding enthalpy to the histone particle, as well as local DNA bending and twisting energy.  If we assume only
non-sequence specific interactions and uniform extension energies ($\ve=\ve_{0}$), $f(y_{i+1},y_{i}) =
f(y_{i+1}-y_{i})$. The integration limits in (\ref{Z0}) reflect the particles' impenetrability and renders
$Z(N,L)$ convolutions of the functions $f(y_{i+1}-y_{i})$.  Upon using Laplace transforms in the variable $L_{0}$,

\begin{equation}
Z(N,L) = \int_{\gamma-i\infty}^{\gamma+i\infty} 
\tilde{f}^{N}(s)e^{sL_{0}}{ds \over 2\pi i},
\label{ZL}
\end{equation}

\noindent where $\tilde{f}(s)$ is the Laplace transform of 
$f$ and $\gamma \in {\cal R}$ is greater than the real parts of
all singularities of the integrand.  From (\ref{ZL}),  one readily finds
thermodynamic quantities such as moments of the 
particle lengths $\ell_{i}$,

\begin{equation}
\langle \ell_{i}^{n}\rangle = {(-1)^{n}\over Z}\oint 
\tilde{f}^{N-1}(s){\partial^{n}\tilde{f}(s) \over 
\partial \ve_{0}^{n}} e^{sL_{0}} {ds \over 2\pi i},
\label{ELL}
\end{equation}

\noindent positions $\langle y_{i}^{n}\rangle$, and 
particle separations $\langle(y_{i+1}-y_{i})^{n}\rangle$.



Although these quantities are readily computed using specific
functions $\tilde{f}(s)$, the resulting sums typically involve numerical manipulation of
extremely large numbers, especially for large $N$. Thus, it is 
also useful to derive from (\ref{ZL}), using
steepest descents, the leading order large $N$  asymptotic approximation
$Z(N\rightarrow\infty, N/L = \rho) \sim 
e^{N(s^{*}(1/\rho-a) + \ln \tilde{f}(s^{*}))}$,
where $s^{*}$ is the saddle point defined by largest real root of 

\begin{equation}
{1\over \tilde{f}(s^{*})}{\partial \tilde{f}(s)\over \partial s} \bigg\vert_{s=s^{*}}+ 
\left({1\over \rho}-a\right) = 0.
\label{SADDLE}
\end{equation}

\noindent Thermodynamic limits of, 
for example,  the moments of the mean particle lengths (\ref{ELL}) become

\begin{equation}
\displaystyle \langle \ell_{i}^{n}\rangle \sim {(-1)^{n} \over 
\tilde{f}(s)}{\partial^{n} \tilde{f}(s) \over \partial 
\ve_{0}^{n}}\Bigg\vert_{s=s^{*}}.
\label{ASYMPSTATS}
\end{equation}

In the histone winding problem (Fig. \ref{HISTONE}) the unit of length will be a single nucleic acid
base pair (bp). The total length $\ell_{i}$ of each particle  corresponds to the arc-length of
polymer that is wrapped around, and in direct contact with, a histone particle.  The hard core
cut-off $a$ depends on details of the three-dimensional arrangement of adjacent histones, and is
roughly (or slightly smaller than) the diameter of a histone particle. Only for very specific phased
orientations of canted histones along DNA can the histones be spaced less than about $a \approx
20$bp \cite{SINDEN,XRAY1,WIDOM2,RUDNICK}.  The finite width of the histone limits ``winding-in''
lengths $\ell_{i}$ to either the distance to the start of an adjacent particle, $x_{i+1}-x_{i}-a =
y_{i+1}-y_{i}$, or to $w \approx 146$bp, the maximum winding length corresponding to 1.7 loops
around a histone particle. For example, particles zero and one in  Fig. \ref{HISTONE}a have only
one base pair of contact ($\ell_{0}=\ell_{1}=0$), particles two and three are fully wound in
($\ell_{2}=\ell_{3}\approx 146$), while particle four is partially wound in ($0<\ell_{4}<w$).  Upon
imposing the physical limits on the particle lengths,

\begin{equation}
\ell_{i}^{*} = \Bigg\{\begin{array}{lr}
x_{i+1}-x_{i}-a \quad\quad & x_{i+1}-x_{i}-a < w \\[13pt]
w \quad\quad & x_{i+1}-x_{i}-a > w,
\end{array}
\label{LSTARW}
\end{equation}

\noindent in the integration limit in (\ref{F}),  we find 

\begin{equation}
\tilde{f}(s) =
{1-e^{-(s+\ve_{0})w}\over s(s+\ve_{0})}.
\label{FS}
\end{equation}

\noindent With this form of $\tilde{f}(s)$,
the integrand in (\ref{ZL}) appears to have a pole only at $s=0$; however,
an implicit constraint on the number of particles of total length $a+\ell$ that
can fit into $L$ will ``induce'' poles at $s=-\ve_{0}$.  This is seen by
expanding $\tilde{f}^{N}(s)$ in the integrand of (\ref{ZL}),

\begin{equation}
Z(N,L) =\sum_{k=0}^{N} 
a_{k}
\int_{\gamma-i\infty}^{\gamma+i\infty} 
{e^{-k\ve_{0}w} e^{s(L_{0}-kw)} \over s^{N}(s+\ve_{0})^{N}} {ds \over 2\pi i},
\label{ZEXPAND}
\end{equation}

\noindent where $a_{k}\equiv (-1)^{k}{N \choose k}$. 
For $\gamma > \mbox{max}\{0,-\ve_{0}\}$  and
$(L-Na-kw) > 0$, we close the contour in the left-hand $s$-plane. For $L$
and $k$ such that  $(L-Na-kw) < 0$, convergence demands that we
close the contour in the right $s$-half-plane.  Since there are no poles to
the right of $\gamma$, terms with $(L-Na-kw) < 0$ correspond to
configurations with more particles in $L$ than is possible, and do
not contribute to the partition function.  Therefore, we need only sum
(\ref{ZEXPAND}) to $k=k^{*} =\mbox{min}\{ \mbox{int}\left[(L-Na)/w\right],
N\}$. Expanding all terms and explicitly evaluating the residues at
$s=0, -\ve_{0}$, we obtain the exact expression,

\begin{equation}
\begin{array}{l}
\displaystyle Z(N\geq 1,L)= {(-1)^{N}N\over N!\ve_{0}^{2N-1}}\sum_{k,p=0}^{k^{*},N-1}
a_{k}b_{p}(L-Na-kw)^{p} \\[12pt]
\hspace{2.2cm} \displaystyle \times\ve_{0}^{p}\left[e^{-\ve_{0}(L-Na)}-(-1)^{p}e^{-k\ve_{0}w}\right],
\end{array}
\label{ZEXACT}
\end{equation}


\noindent where $b_{p}\equiv (2(N-1)-p)!/(p!(N-1-p)!)$.  Numerically, the asymptotic approximation
(\ref{ASYMPSTATS}) is accurate to within $\sim$3\% of the exact result 
(\ref{ELL}) provided $N\gtrsim 10$.  

\begin{figure}[htb]
\begin{center}
\includegraphics[height=3.2in]{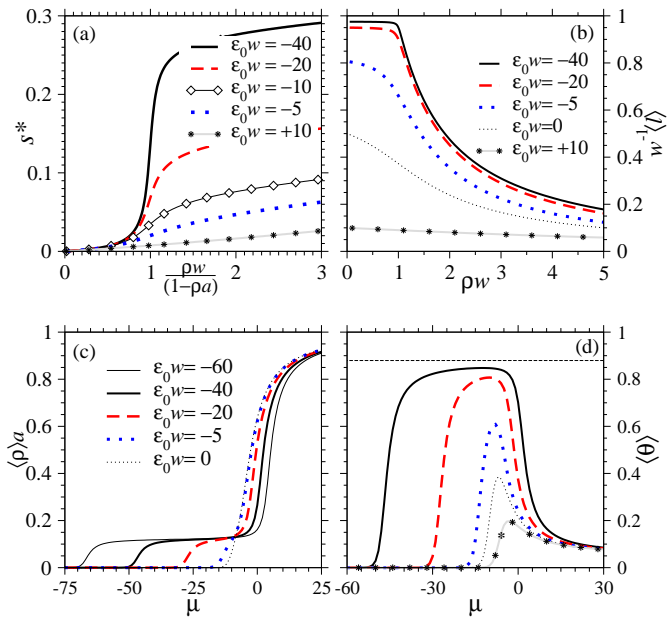}
\end{center}
\caption{Thermodynamic properties for $a=20,w=146$, and various $\ve_{0}$.
(a) The root $s^{*}$ of (\ref{SADDLE}) that defines 
$Z(N\rightarrow \infty, N/L = \rho)$ and determines
the next-neighbor correlation function $g^{(2)}(x_{i+1}-x_{i}\vert N) =
s^{*}e^{-s^{*}(x-a)}$. (b) Winding-in lengths $w^{-1}\langle\ell\rangle$
as a function reduced density $\rho w$. 
(c) Number density as a function of chemical potential $\mu$
for various binding affinities $\ve_{0}w$. An intermediate density 
plateau at $\langle\rho\rangle a \approx a/(w+a)$ 
arises for high affinity histones due to close packing of 
fully wound-in particles of length $w+a$.
(d) Mean coverage as a function of $\mu$. Curves correspond to the 
legend used in (b).  Maxima arise 
for wound-in, high affinity particles spaced 
an average distance $w+a$ apart. The high compression ($\mu\rightarrow \infty$)
limit forces unwinding but packs histones at intervals of the linker
length $a$ for a coverage fraction $\langle\theta\rangle \approx 1/a$.}
\label{figw}
\end{figure}

Fig. \ref{figw}a plots the saddle solution $s^{*}$ found from (\ref{SADDLE}).  In the fixed $N$
ensemble, the probability distribution for particle $i+1$ to be at position $x_{i+1}$ given that
particle $i$ is at position $x_{i}$ is readily computed in the asymptotic limit ($L\gg \Lambda \gg
a$), $g^{(2)}(x_{i+1}-x_{i}\vert N) \sim Z(N,L-x)/\int_{a}^{\Lambda} Z(N,L-x)dx =
s^{*}e^{-s^{*}(x-a)}$. This result implies that the adjacent particle is statistically confined
to within $x \lesssim 1/s^{*}$ \cite{PELITI}.  At low number densities, adjacent histones are spaced
far apart and $s^{*}\sim \rho/(1-\rho a)$.  For attractive lengthening interactions ($\ve_{0}\ll
0$), a sharp increase in $s^{*}$ occurs near $\rho \sim 1/(w+a)$ signaling a partial confinement of
hard rod particles of roughly size $w+a$.  At extremely high densities, $\rho \sim a^{-1}$,
particles are compressed at the expense of unwinding, and $s^{*}$ increases further as $s^{*} \sim
2\rho/(1-\rho a)-\ve_{0}/2$. Fig. \ref{figw}b shows the mean winding-in length (normalized by $w$),
found from (\ref{ASYMPSTATS}). At low densities and strong attractive binding, the maximal winding
in length $w^{-1}\langle \ell \rangle \sim 1$ is approached, while at high densities, the winding-in
length is restricted by nearest neighbors and $\langle \ell \rangle \approx 1/\rho$.

Histone-DNA affinity and competition experiments however, are performed by exposing
DNA to fixed bulk histone concentrations \cite{WIDOM2,CROTHERS}. When the mean bound histone
number is determined by the bulk histone chemical potential (thus
not necessarily large), we employ the grand partition function found
using the exact expression (\ref{ZL}) or (\ref{ZEXACT}) in $\Xi(\lambda,L)
\equiv \sum_{N=1}^{N^{*}}\lambda^{N}Z(N,L)$. The fugacity $\lambda \equiv
e^{\mu-\ve_{0}}$ takes into account the bulk histone chemical potential,
and the binding energy $\ve_{0}$ of a single base pair. $N^{*}= \mbox{int}\{L/a\}$ is
the maximum number of particles that can fit into length $L$.  In the $L/a
= \infty$, $N^{*}=\infty$ limit, we find 

\begin{equation}
\Xi(\lambda,L\rightarrow \infty) = \oint {\lambda e^{-sa}\tilde{f}(s) e^{sL} \over 
1-\lambda e^{-sa}\tilde{f}(s)} {ds \over 2\pi i},
\label{XIDEF}
\end{equation}

\noindent which can be evaluated using (\ref{FS})
to yield $\Xi(\lambda, L\rightarrow \infty)=\langle\rho\rangle e^{s_{+}L}$,
where the mean density $\langle\rho\rangle \equiv L^{-1}\partial_{\mu}\ln \Xi$
is explicitly

\begin{equation}
\langle\rho\rangle  =
{s_{+}(s_{+}+\ve_{0})
\over 2s_{+}+\ve_{0}+\lambda e^{-s_{+}a}(a-(w+a)e^{-(s_{+}+\ve_{0})w})},
\label{RHOW}
\end{equation}

\noindent $s_{+}$ being the largest real root of
$1-\lambda \tilde{f}(s_{+})e^{-s_{+}a}=0$.
The mean fraction of DNA base pairs covered by contacts with histones is then found 
from

\begin{equation} 
\langle \theta\rangle =-{1\over L}{\partial \ln \Xi \over
\partial \ve_{0}}+\langle\rho\rangle.
\label{COVERAGEW}
\end{equation}

The mean density and coverage are plotted in Figs. \ref{figw}c,d. For $\mu\rightarrow -\infty$,
$\langle \rho\rangle \sim s_{+} \sim e^{-\ve_{0}+\mu}(1-e^{-\ve_{0} w})/\ve_{0}\rightarrow 0$ since
this limit corresponds to infinitely dilute bulk histone concentration. Densities of adsorbed
particles increase with bulk histone chemical potential.  For higher affinity histones, these
increases occur earlier and plateau at a value corresponding to close packing of fully wound-in
particles, or $\langle \rho\rangle a \approx a/(w+a) \approx 0.12$.  As $\mu$ and the density
further increase, particles compress and unwind each other until the density approaches maximal
packing at $\langle\rho(\mu\rightarrow\infty)\rangle \sim 1/a-2/(\mu a) + O(\mu^{-2}\ln\mu)$.  This
crossover behavior is seen only for high affinity ($\ve_{0}\ll 0$) histones. The transition from monotonic to
the two-stage density behavior may be observable by tuning the nonspecific histone binding energies
$\ve_{0}$ by {\it e.g.} varying ionic strength \cite{WIDOM2,NETZ}.

DNA accessibility (and hence occlusion by histone particles) and positioning is thought to be a major determinant
of nucleic acid processing by other regulatory proteins \cite{SINDEN,WIDOM,UCLA,WIDOM2}.  The total coverage
$\langle \theta\rangle$ defined as the mean fraction of base pairs in contact with histones, is shown in Fig. 
\ref{figw}d. For $\mu\rightarrow -\infty$, there are few particles present to adsorb onto the DNA substrate and
$\langle\theta\rangle \sim e^{-\ve_{0}+\mu}\rightarrow 0$.  An increase in mean coverage results from increasing
$\mu$ and the number of bound histones.  The theoretical maximum coverage $w/(w+a) \approx 0.88$ (thin dashed
line) corresponds to close packing of fully wound-in particles and is approached only for high affinity histones.
The minimum {\it uncovered} fraction $a/(w+a)$ results from the linker DNA of minimum length $a\sim 20$ joining
two adjacent nucleosomes. If $\mu$ is further increased, the particles squeeze on each other until they unwrap to
the point where only a single base pair remains in contact for each histone.  In this limit, the mean coverage
$\langle \theta(\mu\rightarrow \infty)\rangle \sim 1/a+(1-2/a)/\mu + O(\mu^{-2}\ln\mu) \approx 0.05$ while the
histones are spaced at their maximal densities $\langle \rho \rangle \approx 1/a$.  The {\it variance} in
coverage, $\mbox{var}(\theta)= L^{-1}\partial (L^{-1}\partial\Xi/\partial \ve_{0} +\Xi)/\partial \ve_{0}$, gives
a standard deviation in coverage proportional to the mean coverage, $\sqrt{\mbox{var}(\theta)} =
\sqrt{2}\langle\theta\rangle$.  Despite the high coverage at intermediate chemical potentials, {\it fluctuations}
in this regime are also maximal, suggesting a dynamically controlled DNA accessibility mediated by
histone-histone interactions.  

\begin{figure}[htb]
\begin{center}
\includegraphics[height=1.9in]{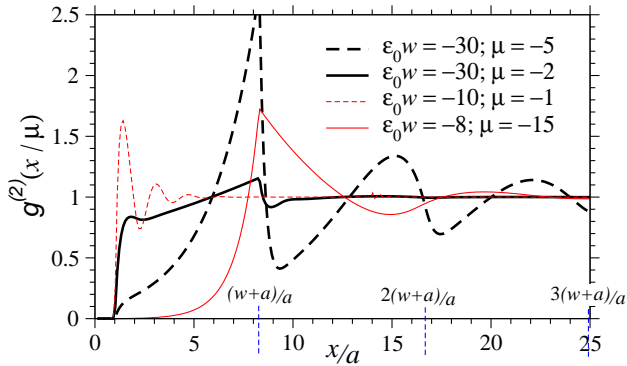}
\end{center}
\caption{The correlation $g^{(2)}(x\vert \mu)$ exhibits properties of 
both length scales $a$ and $w$ depending on $\ve_{0}$ and $\mu$. For high affinity 
and low densities
({\it e.g.} $\ve_{0}w=-30, \mu=-5$), the density distribution is similar to that 
of a hard rod Tonks gas  of length $a+w$.
Upon increasing the density ($\ve_{0}w=-30, \mu =-2$), 
features associated with both length scales arise as partial unwinding occurs,
exposing  the hard core repulsion of size $a$. Conditions under which particles weakly wind
in ($\ve_{0}w=-10$) exhibit behavior attributed to hard rods of length $a$. However, even 
at lower affinities ($\ve_{0}w=-8$), behavior approximating that of rods of length $w+a$ can be recovered if 
densities are made sufficiently low ($\mu =-15$).}
\label{figwcorr}
\end{figure}

In the grand ensemble, the correlation function analogous to $g^{(2)}(x_{i+1}-x_{i}\vert N)$ is
$\langle\rho\rangle g^{(2)}(x\vert \mu)\equiv \Xi(\lambda,x)\Xi(\lambda,L-x)/\Xi(\lambda,L)$ which
describes the probability distribution that given a particle at the origin, {\it any} other histone
exists at $x$.  When $x \not\gg 1/\langle\rho\rangle$ and $N^{*}< \infty$, the exact expression
(\ref{ZEXACT}) must be used to compute $\Xi(\lambda,x)$. Fig. \ref{figwcorr} shows
$g^{(2)}(x\vert\mu)$ computed for various values of $\ve_{0},\mu$.  
Finite particle size anticorrelations give rise 
to oscillations at two length scales 
(see Fig. \ref{figwcorr} caption) depending on affinity  $\ve_{0}$ and density
$\langle\rho(\mu)\rangle$.


The proposed model considers only the adjacent histone exclusion interactions mediated by
nonoverlapping footprints and neglects sequence specific and nucleosome-nucleosome interactions
arising in compact, 3D chromatin structure.  Nonetheless, our theory can be solved exactly with
uniform wrapping energies $\ve_{0}$ to give reasonable results for winding-in lengths
$\langle\ell\rangle$ and histone-histone correlations $g^{(2)}(x\vert N)$ and $g^{(2)}(x\vert \mu)$.
The model predicts a two-state adsorption process and a maximum in the DNA coverage fraction
$\langle\theta\rangle$ as a function of bulk histone concentration and binding affinity.  However,
fluctuations in the coverage are also concurrently maximal during the peak in mean coverage,
indicating that thermal effects can nonetheless provide dynamic access to highly covered DNA. These
predictions  can be tested experimentally by varying bulk histone coverage and ionic strength
(although this would also affect DNA bendability and $\ve_{0}$)\cite{WIDOM2,NETZ}, provided short
enough DNA segments are used such that large scale 3D structures do not arise and
non-nearest-neighbor interactions remain irrelevant.

The model can be readily generalized to include the effects of linker DNA twist, externally applied
tension \cite{MARKO}, and relative histone orientation \cite{RUDNICK} on the cut-off $a$. 
Specific sequences, and their effects on local bendability, twistability, and histone affinity has
also been found to be important {\it in vitro} \cite{WIDOM2} and can be treated within a similar
framework, although completely generalizing our model to include specific sequences
\cite{WIDOM2,CROTHERS} (spatial dependence of $\ve(x)$), would require computational approaches.

Finally, we can extend the Tonks-Takahashi model by removing the maximal winding length
constraint ($w\rightarrow \infty$). The resulting model can then be applied to problems of 1D
nucleation and polymerization.  Particles would correspond to a nucleation domains that cannot
coalesce due to perhaps growth asymmetry (as in actin filaments) or incommensurability with the
underlying substrate \cite{VILFAN,HIPPEL}.  The ``winding-in'' or polymerization length $\ell$ is
now the degree of polymerization in that domain.  Since $\ell_{i}$ are limited only by the position
of adjacent domains, $\ell_{i}^{*} = x_{i+1}-x_{i}-a$ and $\tilde{f}(s) = [s(s+\ve_{0})]^{-1}$. 
Results analogous to those obtained for the histone problem are found by setting $w=\infty$ in
(\ref{RHOW}-\ref{COVERAGEW}). The qualitative behavior of $\ell(\ve_{0})$ can be found in closed
form in the $N=\infty, a=0$ limit: $\rho^{n}\langle \ell_{i}^{n}\rangle \sim 2^{n}n!/(z+2 +
\sqrt{z^{2}+4})^{n}$, where  $z\equiv \ve_{0}(1-\rho a)/\rho$.  This result also implies
$\langle\ell_{i}^{2}\rangle-\langle\ell_{i}\rangle^{2} = \langle\ell_{i}\rangle^{2}$.  The behaviors
for $w = \infty$  are qualitatively different from those shown in Fig. \ref{figw}. Namely, the
adsorption isotherm is monotonic and the coverage has a {\it minimum} at intermediate $\mu$ when
$\ve_{0} <0$ and vanishes for $\mu \rightarrow -\infty$ if $\ve_{0}>0$.  Such differences highlight
the importance of the additional length scale $w$ unique to histone binding.

TC thanks M. D'Orsogna for comments and the NSF for support through grant DMS-0206733.


\end{document}